\documentclass[floatfix,aps,prd,showpacs,amsmath,nofootinbib,preprintnumbers,twocolumn]{revtex4}

\usepackage{graphicx}
\usepackage{bm}

\newcommand{\figurewidth}{3.4in}

\def\half{{1\over 2}}

\def\half{{1\over 2}}
\def\({\left(}
\def\){\right)}
\def\[{\left[}
\def\]{\right]}

\def\e{\begin{equation}}
\def\q{\end{equation}}
\def\m{\begin{eqnarray}}
\def\n{\end{eqnarray}}

\begin{document}

\title{Large Non-Gaussianity Implication for Curvaton Scenario}

\author{Qing-Guo Huang}\email{huangqg@kias.re.kr}
\affiliation{School of physics, Korea Institute for Advanced Study,
207-43, Cheongryangri-Dong, Dongdaemun-Gu, Seoul 130-722, Korea}

\date{\today}

\begin{abstract}

We argue that the typical energy density of a light scalar field
should not be less than $H^4$ in the inflationary Universe. This
requirement implies that the non-Gaussianity parameter $f_{NL}$ is
typically bounded by the tensor-scalar ratio $r$ from above, namely
$f_{NL}\lesssim 518\cdot r^{1\over 4}$. If $f_{NL}=10^2$, inflation
occurred around the GUT scale.

\end{abstract}


\maketitle


\section{Introduction}

Inflation model \cite{Guth:1980zm,Albrecht:1982wi,Linde:1982zj}
provides an elegant mechanism to solve the horizon, flatness and
primordial monopole problem due to a quasi-exponential expansion of
the universe before the hot big bang. The temperature anisotropies
in cosmic microwave background radiation (CMBR) and the large-scale
structure of the Universe are seeded by the primordial quantum
fluctuations during inflation \cite{Mukhanov:1990me,Lyth:1998xn}.
Since the density perturbation is roughly $10^{-5}$, it is good
enough to apply the linear perturbation theory to calculate the
quantum fluctuations during inflationary epoch. Within this
approach, the Fourier components of fluctuations are uncorrelated
and their distribution is Gaussian. That is why the non-Gaussianity
from the simplest inflation models is very small ($|f_{NL}|<1$). For
useful discussions on non-Gaussianity see
\cite{Salopek:1990jq,Salopek:1990re,Falk:1992sf,Gangui:1993tt,Acquaviva:2002ud,Maldacena:2002vr},
and for a nice review see \cite{Bartolo:2004if}.

The non-Gaussian perturbation is governed by the n-point correlation
function of the curvature perturbation \e \langle
\Phi({\bf{k}_1})\Phi({\bf{k}_2})\cdot\cdot\cdot\Phi({\bf{k}_n})\rangle=(2\pi)^3\delta^3(\sum_{i=1}^{n}{\bf{k}_i})F_n(k_1,k_2,...,k_n),\q
where $\Phi({\bf{k}})$ is the Fourier mode of Bardeen's curvature
perturbation. The leading non-Gaussian features are known as the
bispectrum (three-point function) and trispectrum (four-point
function), with their sizes conventionally denoted as $f_{NL}$ and
$\tau_{NL}$ respectively. The non-linearity parameter $f_{NL}$
defined in \cite{Komatsu:2001rj} takes the form \e \Phi({\bf
x})=\Phi_L({\bf x})+f_{NL}[\Phi_L^2({\bf x})-\langle\Phi_L^2({\bf
x})\rangle], \q where $\Phi_L$ denotes the linear Gaussian part of
the perturbation in real space. The non-Gaussianity parameter
$f_{NL}$ characterizes the amplitude of the primordial non-Gaussian
perturbations. The value of $\Phi_L$ is roughly $10^{-5}$. If
$f_{NL}=10^2$, the distribution of $\Phi$ is still consistent with a
Gaussian distribution to $0.1\%$. It is not easy to detect a small
non-Gaussianity in the experiments. The shape of non-Gaussianity is
also very important. If $F(k_1,k_2,k_3)$ is large for the
configurations in which $k_1\ll k_2,k_3$, it is called local,
``squeezed" type; if $F(k_1,k_2,k_3)$ is large for the
configurations in which $k_1\sim k_2\sim k_3$, it is called
non-local, ``equilateral" type.

Recently a large primordial local-type non-Gaussianity was reported
to be marginally detected from the data of WMAP3 in
\cite{Yadav:2007yy}: \e 27<f_{NL}^{local}<147\q at $95\%$ C.L., with
a central value of $f_{NL}=87$. See also \cite{Komatsu:kitpc}. In
\cite{Komatsu:2008hk} WMAP group recently reported their new result
as \e -9<f_{NL}^{local}<111. \q If a large local-type
non-Gaussianity is confirmed by the future cosmological observations
and data analysis at high confidence level, it strongly implies that
the physics of the early Universe is more complicated than the
simple single-field slow-roll inflation.

There are several mechanisms to generate a large local-type
non-Gaussianity:\\
(1) a non-linear relation between inflaton and curvature perturbations \cite{Gangui:1993tt}; \\
(2) curvaton scenario \cite{Linde:1996gt,Enqvist:2001zp,Lyth:2001nq,Moroi:2001ct,Lyth:2002my,Bartolo:2003jx}; \\
(3) Ekpyrotic model \cite{Creminelli:2007aq,Buchbinder:2007at,Koyama:2007if,Lehners:2007wc}.\\
In this paper we focus on curvaton scenario. As we know, there might
be many light scalar fields in supersymmetric theory or string
theory. These light scalar fields which are subdominant during
inflation are called curvaton. The energy density during inflation
is still dominated by the the potential of inflaton. In the usual
inflation model the fluctuations of inflaton dominate the curvature
perturbations. The energy density and the perturbations caused by
these light scalar fields can be ignored during inflation. However
it is possible that the fluctuation of curvaton becomes relevant and
causes a large local-type non-Gaussianity when its energy density is
a significant fraction of the total energy after the end of
inflation. In curvaton scenario the perturbations from the inflaton
field are considered to be negligible. A large local-type
non-Gaussianity may shed light on these light scalar fields.

The non-Gaussianity produced by curvaton is in inverse proportion to
the fraction of curvaton energy density in the energy budget at the
epoch of curvaton decay. The non-Gaussianity increases as the
curvaton energy density decreases. However the energy density for a
homogeneous scalar field in the quasi-de Sitter space is typically
is given by $H_*^4$. Therefore the non-Gaussianity has an upper
bound \e f_{NL}\lesssim 518\cdot r^{1\over 4}, \q where $r$ is the
tensor-scalar ratio. Present bound on the tensor-scalar ration
\cite{Komatsu:2008hk} is $r<0.20$ and then $f_{NL}\lesssim 346$.

Our paper is organized as follows. We review curvaton scenario and
the curvature perturbation generated by curvaton in Sec. 2. In Sec.
3, we argue that the energy density of curvaton is typically
estimated as $H_*^4$ during inflation and then the non-Gaussianity
parameter $f_{NL}$ is bounded by the tensor-scalar ratio from above.
In Sec. 4, we point out the challenge for getting a red-tilted
primordial power spectrum in curvaton scenario. Section 5 contains
some discussions including how to distinguish the curvaton scenario
from Ekpyrotic scenario.

\section{Non-Gaussianity in curvaton scenario}

In curvaton scenario the dynamics of inflation in the early Universe
is dominated by inflaton and the energy density of curvaton is
negligible during inflation. The lagrangian of a curvaton field
$\sigma$ in an unperturbed Friedmann-Robertson-Walker (FRW)
spacetime is given by \e {\cal L}_\sigma=\half {\dot
\sigma}^2-\half(\nabla_p \sigma)^2-V(\sigma), \q where the subscript
``$p$" denotes the derivative with respect to the physical
coordinate. For simplicity, we consider a potential of curvaton as
follows \e V(\sigma)={1\over 2} m^2\sigma^2. \q The curvaton field
is supposed to be an almost free field with a small effective mass
compared to the Hubble scale $H$ during inflation. Effectively we
define a parameter \e \eta_{\sigma\sigma}\equiv {1\over
3H^2}{d^2V(\sigma)\over d\sigma^2} \q which is much smaller than 1.
The equation of motion for a homogeneous curvaton field takes the
form \e \ddot \sigma+3H\dot \sigma=-m^2\sigma. \label{eom} \q Since
$m\ll H$, the friction term $3H\dot \sigma$ is dominant and the
value of curvaton is roughly a constant at the inflationary epoch.
It is denotes as $\sigma=\sigma_*$.

The amplitude of the quantum fluctuation of curvaton field in a
quasi-de Sitter space is given by \e \delta \sigma={H_*\over 2\pi}.
\q The spectrum of the fractional perturbations caused by curvaton
\cite{Lyth:2001nq} is \e P_{\delta\sigma/\sigma}^{1\over 2}\simeq
{\delta \sigma\over \sigma}={1\over 2\pi}{H_*\over \sigma_*},\q
where $*$ denotes the epoch of horizon exits $k_*=a_*H_*$ during
inflation. The quantum fluctuation of curvaton field is frozen at
horizon exit to a classical perturbation with a flat spectrum. Since
the curvaton energy density is subdominant in this epoch, its
fluctuations are initially taken as isocurvature/entropy
fluctuations. But the curvature perturbation is sourced on large
scales \cite{Bartolo:2004if}.

After the end of inflation the inflaton energy density is converted
into radiations and then the Hubble parameter decreases as $a^{-2}$.
The curvaton field will remain approximately constant $\sigma_*$
until $H\sim m$. At this epoch the curvaton starts to oscillate
harmonically about $\sigma=0$. During the stage of oscillating the
curvaton energy density goes like $\rho_\sigma\propto a^{-3}$ which
increases with respect to the energy density of radiation
$\rho_R\propto a^{-4}$. When the Hubble parameter goes to the same
order of the curvaton decay rate $\Gamma$, the curvaton energy is
converted into radiations. Finally, before primordial
nucleosynthesis, the curvaton field is supposed to completely decay
into thermalized radiation, thus the perturbations in the curvaton
field are converted into curvature perturbations and become the
final adiabatic perturbations which seed the matter and radiation
density fluctuations observed in the Universe. We illustrate the
evolution of one curvaton perturbation mode and Hubble radius in
Fig. 1.
\begin{figure}[ht]
\centerline{\includegraphics[width=\figurewidth]{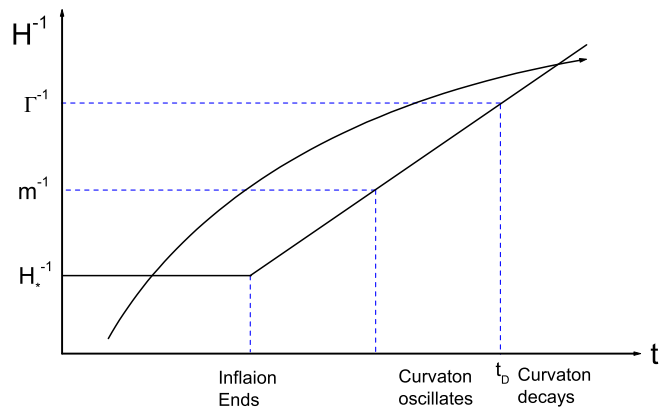}}
\caption{Evolution of the perturbation mode and the Hubble radius.}
\end{figure}

After the curvaton decay the total curvature perturbation will
remain constant on superhorizon scales at a value which is fixed at
the epoch of curvaton decay. The sudden-decay approximation is exact
if the curvaton completely dominates the energy density before it
decays. Going beyond the sudden-decay approximation, a numerical
calculation in \cite{Malik:2002jb} implies that the spectrum of the
curvature perturbations caused by curvaton is proportional to the
fraction of curvaton energy density in the energy budget at the
epoch of curvaton decay \e \Omega_{\sigma,D}=\({\rho_{\sigma}\over
\rho_{tot}}\)_D.\q A precise form of the spectrum in curvaton
sceanrio is given in \cite{Lyth:2002my} as follows \e
P_{\zeta_\sigma}^{1\over 2}={2\over
3}\Omega_{\sigma,D}P_{\delta\sigma/\sigma}^{1\over 2}={1\over
3\pi}\Omega_{\sigma,D}{H_*\over \sigma_*}. \label{zs} \q In curvaton
scenario the curvature perturbation is dominated by curvaton
perturbation and then the spectral index in the curvaton model takes
the form \e n_s=1+2\eta_{\sigma\sigma}-2\epsilon_H, \q where
$\epsilon_H$ is the slow-roll parameter which is defined as \e
\epsilon_H\equiv-{\dot H_*\over H_*^2}.\q In
\cite{Lyth:2002my,Lyth:2006gd} the non-Gaussianity parameters
corresponding to bispectrum \footnote{A precise form in
\cite{Bartolo:2004if} is $f_{NL}={2\over 3}-{5\over
6}\Omega_{\sigma,D}+{5\over 4\Omega_{\sigma,D}}$.} and trispectrum
are respectively given by  \e f_{NL}\simeq {5\over
4\Omega_{\sigma,D}},\label{fnl}\q and \e \tau_{NL}={36\over
25}f_{NL}^2. \q In sudden-decay approximation, Eq.(\ref{fnl}) should
be replaced by $5/(3\Omega_{\sigma,D})$. A large non-Gaussianity is
obtained if $\Omega_{\sigma,D}\ll 1$. In this case the Hubble
parameter is always dominated by the radiation energy density before
curvaton decays. Assume the scale factor is $a(t_o)=1$ at the moment
when curvaton start to oscillate. The energy density of curvaton and
radiation are $\rho_{\sigma}(t_o)=\half m^2\sigma_*^2$ and
$\rho_R(t_o)=3M_p^2m^2$ respectively. Since the radiation energy
density goes like $a^{-4}$, $\rho_R(t_D)\simeq 3M_p^2\Gamma^2\simeq
\rho_R(t_o)a^{-4}(t_D)$ and thus $a(t_D)\simeq (m/\Gamma)^\half$.
Therefore \e \Omega_{\sigma,D}\simeq {\rho_\sigma(t_D)\over
\rho_R(t_D)}={\sigma_*^2\over 6M_p^2}\({m\over \Gamma}\)^\half.
\label{dom}\q This result is also given in \cite{Lyth:2001nq}.
Combing (\ref{fnl}) with (\ref{dom}), we get \e f_{NL}={15\over
2}{M_p^2\over \sigma_*^2}\({\Gamma\over m}\)^\half.\label{fnls}\q
Since $\Gamma<m$, a large non-Gaussianity is obtained only when
$\sigma_*\ll M_p$. It is reasonable to require that the VEV of
curvaton is less than Planck scale and a large $f_{NL}$ is expected.

The WMAP normalization \cite{Komatsu:2008hk} is \e
P_{\zeta}=2.457\times 10^{-9}. \label{coben}\q Using eq. (\ref{zs})
and (\ref{fnl}), the value of curvaton during inflation is related
to the Hubble parameter by \e \sigma_*=2.68\times 10^3 {1\over
f_{NL}}H_*.\label{ssh} \q The non-Gaussianity parameter $f_{NL}$
should be smaller than $2.68\times 10^3$; otherwise, the quantum
fluctuation of curvaton is greater than its VEV and the previous
semi-classical description breaks down. Substituting (\ref{ssh})
into (\ref{fnls}), we find \e f_{NL}\simeq 10^6{H_*^2\over
M_p^2}\({m\over \Gamma}\)^\half.\label{fnlg}\q The limit of the
primordial gravitational wave perturbation implies that
$H_*<10^{-4}M_p$, and then the mass of curvaton must be much larger
than its decay rate for getting a large non-Gaussianity.

Tensor/gravitational wave perturbation encodes a very important
information about inflation. Its amplitude only depends on inflation
scale \e P_T={H_*^2/M_p^2\over \pi^2/2}. \label{ts}\q For
convenience, we define a new parameter named the tensor-scalar ratio
$r$ as \e r={P_T\over P_\zeta}. \q If the curvature perturbation is
dominated by inflaton field we have $r=16\epsilon_H$ for slow-roll
inflation. In curvaton scenario the density perturbation is
dominated by curvaton and thus \e r<16\epsilon_H.\label{der} \q
According to (\ref{ts}), the Hubble parameter during inflation is
related to the tensor-scalar ratio by \e H_*=10^{-4}r^\half
M_p=2.44\times 10^{14} r^\half \hbox{GeV}. \label{hrm}\q Now eq.
(\ref{fnlg}) becomes \e f_{NL}=10^{-2}r\({m\over \Gamma}\)^\half.\q
WMAP5 gives an upper bound on the tensor-scalar ratio $r<0.20$
\cite{Komatsu:2008hk} and we need $m/\Gamma>2.5\times 10^9$ for
$f_{NL}=10^2$. The bound on the tensor perturbation also provides an
upper bound on the inflation scale \e H_*<1.09\times 10^{14}
\hbox{GeV}.\q We will see that the non-Gaussianity in curvaton
scenario will give a lower bound on $H_*$ in the next section.

\section{Upper bound on non-Gaussianity $f_{NL}$ in curvaton scenario}

The curvaton mass and its decay rate are fixed if the field theory
for the whole system is given. Since the curvaton does not move
during inflation, we cannot use its dynamics to determine the value
of $\sigma_*$. According to Eq.(\ref{fnls}), $f_{NL}$ is large if
$\sigma_*$ is much smaller than the Planck scale. In the
literatures, $\sigma_*$ is not fixed by the theory, but rather
represents an additional free parameter, and then the
non-Gaussianity parameter $f_{NL}$ can be arbitrarily large. This
treatment is reliable at the classical level. However we will see
that the quantum fluctuations of curvaton will significantly affect
the value of curvaton during inflation and an upper bound on the
typical value of $f_{NL}$ is obtained for the curvaton model.

In the inflationary Universe the quantum fluctuations of the
curvaton field freeze out with amplitude $H/(2\pi)$ at the
wavelength $H^{-1}$. In \cite{Linde:1993xx} a gradient energy
density due to these fluctuations is estimated \e (\nabla_p
\sigma)^2\sim H_*^4. \q It is also obtained through calculating the
contribution of these quantum fluctuations to the average value of
the energy momentum tensor. The gradient energy density
characterizes the inhomogeneity of the curvaton and should be
smaller than its homogeneous energy density $\half m^2\sigma_*^2$.
Requiring $m^2\sigma_*^2\gtrsim H_*^4$ yields a lower bound on the
VEV of curvaton field, namely \e \sigma_*\gtrsim{H_*^2\over m}. \q
Since $m< H_*$, $\sigma_*> H_*$ which is the condition for the
validity of the semi-classical description.

Secondly lets take into account the horizon temperature $H_*$ of a
quasi-de Sitter space. Since the curvaton mass is much smaller than
$H_*$, the curvaton ``particles" can be taken as the relativistic
particles. If the initial temperature of curvaton particles is lower
than $H_*$, the quantum fluctuations of the background will
automatically heat their temperature up to $H_*$. So the energy
density of curvaton field should not be smaller than $H_*^4$.

Here we also give the third argument. In curvaton model the curvaton
mass is assumed to be much smaller than the Hubble parameter during
inflation, which means the Compton wavelength is large compared to
the curvature radius of the de Sitter space $H_*^{-1}$. So the
gravitational effects may play a crucial role on the behavior of the
curvaton field. The well-known Bunch-Davies expression
\cite{Bunch:1978yq,Vilenkin:1982wt,Linde:1982uu} for the average
square value of a light scalar field ($m\ll H_*$) in a quasi-de
Sitter space takes the form \e \langle\sigma^2 \rangle={3H_*^4\over
8\pi^2m^2}. \label{bdw}\q An intuitive understanding was given in
\cite{Linde:2005yw}. According to the long-wave quantum fluctuation
of a light scalar field $(m\ll H)$ in inflationary universe, the
behavior of such a light scalar field can be taken as a random walk
\cite{Linde:2005ht}: \e \langle\sigma^2\rangle={H^3\over 4\pi^2}t.
\label{ss} \q On the other hand, a massive scalar field cannot grow
up to arbitrary large vacuum expectation value because it has a
potential. The long wavelength modes of the light scalar field are
in the slow-roll regime and obey the slow-roll equation of motion,
i.e.  \e 3H{d\sigma\over dt}=-{dV(\sigma)\over
d\sigma}=-m^2\sigma.\q Combining these two considerations, in
\cite{Linde:2005yw} Linde and Mukhanov proposed  \e
{d\langle\sigma^2\rangle\over dt}={H^3\over 4\pi^2}-{2m^2\over
3H}\langle\sigma^2\rangle. \label{rw}\q We see that
$\langle\sigma^2\rangle$ stabilizes at the point of $
\langle\sigma^2\rangle={3H^4\over 8\pi^2m^2}$ which is the same as
Eq.(\ref{bdw}). This vacuum expectation value of curvaton mainly
comes from the perturbation mode with wavelength ${H_*^2\over
m^2}H_*^{-1}$ which stretched outside the horizon at the number of
e-folds $H_*^2/m^2$. The inflation models with small total number of
e-folds are artificial and the long stage of inflation is expected
generically \cite{Linde:2003hc,Tegmark:2004qd}. Since the wavelength
is much larger than the Hubble horizon, this fluctuation mode is
frozen to be a classical one and provides a non-zero classical
configuration for curvaton field. The typical value of curvaton
field in such a background is estimated as \e \sigma_*\sim
{H_*^2\over m}. \label{bsg}\q Similar estimation is also adopted in
the literatures, for example \cite{Chun:2004gx}.

The value of $\sigma_*$ is related to the Hubble parameter and
$f_{NL}$ through the WMAP normalization by eq. (\ref{ssh}).
Eq.(\ref{bsg}) reads \e {m\over H_*}\sim {f_{NL}\over 2.68 \times
10^3}. \label{ubh}\q Because the mass of curvaton is smaller than
$H_*$, the non-Gaussianity parameter is less than $2.68\times 10^3$.

The curvaton should decay before neutrino decoupling; otherwise, the
curvature perturbations may be accompanied by a significant
isocurvature neutrino perturbation. The temperature of the universe
at the moment of neutrino decoupling is roughly $T_{nd}=1$ MeV. So
the curvaton decay rate is bounded by the Hubble parameter at the
moment of neutrino decoupling, namely \e \Gamma\gtrsim
\Gamma_0={T_{nd}^2\over M_p}=4.1\times 10^{-25} \hbox{GeV}.
\label{ggt}\q Considering (\ref{fnlg}) and $m<H_*$, we find \e
f_{NL}\lesssim 10^6{H_*^{5\over 2}\over M_p^2\Gamma_0^\half}=
\({H_*\over 2.71\times 10^7 \hbox{GeV}}\)^{5\over 2}.\q Similarly
combining (\ref{fnlg}) and (\ref{ubh}), we have \e f_{NL}\lesssim
\({m\over 4.9\times 10^4 \hbox{GeV}}\)^{5\over 6}.\q For a large
non-Gaussianity $f_{NL}=10^2$, the bounds on the Hubble parameter
and the mass of curvaton become \e H_*\gtrsim 1.71\times 10^8
\hbox{GeV}, \quad \hbox{and}\quad m\gtrsim 1.23\times 10^7
\hbox{GeV}. \q The above constraints on the Hubble parameter and
curvaton mass are not restricted.

In \cite{Lyth:2003dt,Lyth:1995ka} the authors also suggested that
the curvaton decay rate is at least of order $Gm^3\sim m^3/M_p^2$
corresponding to gravitational-strength interactions. So we have \e
\Gamma\gtrsim \Gamma_g\simeq {m^3\over M_p^2}. \label{ggg}\q Now eq.
(\ref{fnlg}) reads \e f_{NL}\lesssim 10^6{H_*^2\over
mM_p}.\label{fhm} \q Including (\ref{ubh}), we find \e
f_{NL}\lesssim \({m\over 3.4\times 10^5 \hbox{GeV}}\)^{1\over 3}.\q
If $f_{NL}=10^2$ the curvaton mass should be larger than $3.4\times
10^{11}$ GeV. Combing (\ref{ubh}), (\ref{fhm}) and (\ref{hrm}), we
find the typical values of $f_{NL}$ and $\tau_{NL}$ are bounded by
the tensor-scalar ratio $r$ from
above: \m f_{NL}&\lesssim&518\cdot r^{1\over 4},\label{fnlr} \\
\tau_{NL}&\lesssim&3.86\cdot 10^5\cdot r^\half. \n Or equivalently,
\e H_*\gtrsim 9\cdot 10^8\cdot f_{NL}^2 \hbox{GeV}.\q Present bound
on the tensor-scalar ratio is $r<0.20$ and then $f_{NL}\lesssim
346$. If $f_{NL}=10^2$, $r\gtrsim 1.4\times 10^{-3}$, $H_*\gtrsim
9\times 10^{12}$ GeV and inflation scale $V^{1/4}\gtrsim 6.2\times
10^{15}$ GeV which implies that inflation happened around GUT scale.
The inequality (\ref{fnlr}) is saturated only when Eq.(\ref{ggg})
are saturated. So the decay rate of curvaton cannot be much larger
than gravitational-strength decay; otherwise the non-Gaussianity
will be small.

The present and next generation of experiments, such as WMAP and
PLANCK, will increase the accuracy to about $|f_{NL}^{cal}|\sim 6$
and $|\tau_{NL}|\sim 560$. If there is really a large local-type
non-Gaussianity, it will be detected at high confidence level in the
near future, and eq. (\ref{fnlr}) provides a restricted constraint
on the curvaton scenario.

\section{Spectral index of the primordial power spectrum}

Spectral index of the primordial power spectrum is another important
quantity for characterizing the physics in the early Universe. WMAP3
\cite{Komatsu:2008hk} prefers a red tilted primordial power spectrum
with $n_s=0.960^{+0.014}_{-0.013}$ for $\Lambda$CDM model, and
$n_s>1$ is disfavored even when gravitational waves ($r<0.2$ at
$95\%$ CL) are included.

The spectral index in the curvaton scenario is given by \e
n_s-1=2\eta_{\sigma\sigma}-2\epsilon_H={2\over 3}{m^2\over
H_*^2}-2\epsilon_H. \label{ns}\q According to eq. (\ref{ubh}), we
see that $\eta_{\sigma\sigma}$ is bounded by the non-Gaussianity
parameter $f_{NL}$ from below, namely \e
\eta_{\sigma\sigma}={m^2\over 3H_*^2}\sim 4.64\cdot 10^{-8}\cdot
f_{NL}^2. \q In curvaton scenario the curvature perturbation is
dominated by curvaton perturbation and then the slow-roll parameter
$\epsilon_H$ has a lower bound (\ref{der}) which says \e
\epsilon_H>{r\over 16}\gtrsim 8.68\cdot 10^{-13}\cdot f_{NL}^4.\q If
$f_{NL}=10^2$, $2\eta_{\sigma\sigma}\sim 9.28\cdot 10^{-4}$ and
$2\epsilon_H\gtrsim 1.74\cdot 10^{-4}$. We see that the
non-Gaussianity cannot give us a useful constraint on the spectral
index.

Usually we assume $m\ll H_*$ and then $\eta_{\sigma\sigma}\simeq 0$.
To get the observed spectral index we need $\epsilon_H\simeq 0.02$
in curvaton scenario. Among so many inflation models, large-field
models, such as chaotic inflation, give roughly the suitable value
of $\epsilon_H$. Large field means the VEV of inflaton is larger
than the Planck scale. However in \cite{Huang:2007gk,Huang:2007qz}
we argued that the VEV of a scalar field should be less than the
Planck scale in a consistent low-energy effective field theory
coupled to gravity. On the other hand, the Lyth bound
\cite{Lyth:1996im} for single-field slow-roll inflation is roughly
given by \e {|\Delta \phi|\over M_p}=\sqrt{2\epsilon_H}\Delta N_e,
\label{lb}\q where $N_e$ is the number of e-fold before the end of
inflation. Requiring $|\Delta \phi|/M_p<1$ yields \e
\epsilon_H<{1\over 2(\Delta N_e)^2}. \q For $\Delta N_e=50$,
$\epsilon_H<2\times 10^{-4}$. So it is reasonable to expect that a
spectral index indistinguishable from 1 is obtained in curvaton
scenario. This is also pointed out in \cite{Dimopoulos:2002kt}. In
inflation model $(n_s-1)$ is mainly due to a negative
$\eta_{\phi\phi}=M_p^2{d^2V(\phi)/d\phi^2\over V(\phi)}$. For the
detail see \cite{Huang:2007qz}.

A possible way to avoid the above bound is to consider multi-field
inflation model. The simplest one is assisted inflation. Since there
is a unique attractor behavior $\phi_1=\phi_2=...=\phi_N$, the
generalized Lyth bound reads \e {|\Delta \phi_i|\over
\Lambda_G}=\sqrt{2\epsilon_H}\Delta N_e, \quad i=1,2,...,N, \q where
$\Lambda_G=M_p/\sqrt{N}$ is the gravity scale for $N$ species
\cite{Huang:2007st,Dvali:2007wp}. In \cite{Huang:2007st} we gave
several examples to support that the variation of each inflaton
should be less than $\Lambda_G$ in the assisted inflation model. If
so, the multi-field inflation cannot help us to release the
constraint on the slow-roll parameter $\epsilon_H$.

Another possibility is chain inflation in string landscape
\cite{Freese:2004vs,Huang:2007ek}. In this scenario, the universe
tunneled rapidly through a series of metastable vacua with different
vacuum energies. Since the total energy density is dominated by the
vacuum energies, we get an inflationary universe. Chain inflation is
not really a slow-roll inflation model and it does not suffer from
the above constraints. In \cite{Huang:2007ek} we have
$\epsilon_H={1\over 3N_e}$. For $N_e=50$, $\epsilon_H=0.0067$,
$n_s=0.987$. Many light scalar fields are expected to emerge in
four-dimensional effective field theory from string theory. Some of
them can be taken as curvatons. The bound on the non-Gaussianity
parameter is $f_{NL}\lesssim 296$. Chain inflation might be generic
in string landscape. So the experiment data can be nicely explained
in string landscape.

Because the curvature perturbation is dominated by curvaton
perturbation in curvaton scenario, we have $r<16\epsilon_H$ and \e
f_{NL}\lesssim 871\cdot \({1\over \Delta N_e}{|\Delta \phi|\over
M_p}\)^\half, \label{fnlp}\q here we take (\ref{fnlr}) and
(\ref{lb}) into account. If $|\Delta\phi|/M_p<1$ for $\Delta
N_e=50$, $f_{NL}\lesssim 123$ which is roughly saturated the
experiment bound.

\section{Discussions}

An unambiguous detection of $f_{NL}>10$ will rule out most of the
existing inflation models. The uncertainty of $f_{NL}^{local}$ from
WMAP will shrink to be 42 for 8 years data, and 38 for 12 years data
\cite{Komatsu:kitpc}. The accuracy of PLANCK will be roughly $6$. If
the local-type non-Gaussianity is of order $10^2$, it will be
detected at high confidence level in the near future.

In this paper we investigate the curvaton scenario in detail and
find that the inflation scale is roughly at GUT scale in order to
get a large local-type non-Gaussianity. Ekpyrotic model can also
provide a large local-type non-Gaussianity. However unlike the
slightly red-tilted gravitational wave spectrum in the
inflation/curvaton model, the gravitational wave spectrum in
Ekpyrotic model is strongly blue and then the amplitude is
exponentially suppressed on all observable scales
\cite{Boyle:2003km}. Gravitational wave perturbation can be used to
distinguish curvaton scenario from Ekpyrotic model.

Here we also want to clarify two points in this paper. One is that
maybe we miss some order-one coefficients in (\ref{bsg}) and
(\ref{ggg}), and then the bound in (\ref{fnlr}) should be modified
to be a little looser or more stringent. But the order of magnitude
of our result can be trusted. The other is that the bound in
(\ref{fnlr}) depends on the WMAP normalization
$P_{\zeta,{WMAP}}=2.457\times 10^{-9}$. If we let the normalization
of the density perturbation free, we find \e f_{NL}\lesssim
\gamma^{-1}\cdot 518\cdot r^{1\over 4}, \q and eq. (\ref{fnlp})
becomes\e f_{NL}\lesssim \gamma^{-1}\cdot 871 \cdot \({1\over \Delta
N_e}{|\Delta \phi|\over M_p}\)^\half, \q where
$\gamma=(P_\zeta/P_{\zeta,{WMAP}})^\half$. Larger the amplitude of
the density perturbation, smaller the non-Gaussianity. If we also
consider $|\Delta \phi|/M_p<1$ for $\Delta N_e=50$, $f_{NL}\lesssim
123\cdot \gamma^{-1}$. The density perturbation cannot be larger
than 10 times of its observed value in our Universe if the
non-Gaussianity parameter is not smaller than 10.

The equilateral-type non-Gaussianity has not been detected. Many
models
\cite{ArkaniHamed:2003uz,Chen:2006nt,ArmendarizPicon:1999rj,Chen:2007gd,Li:2007}
were suggested to generate large equilateral-type non-Gaussianity as
well. Other mechanisms concerning the large non-Gaussianity are
discussed in
\cite{Dvali:2003em,Dvali:2003ar,Matsuda:2007tr,Matsuda:2007ds,Berera:1995ie,Berera:1996nv,Gupta:2002kn}.
Anyway, non-Gaussianity will be an important issue for cosmology and
string theory.

\vspace{.5cm}

\noindent {\bf Acknowledgments}

We would like to thank M.~Sasaki and P.~J.~Yi for useful
discussions.





\begin{thebibliography}{99}
\frenchspacing

\bibitem{Guth:1980zm}
  A.~H.~Guth,
  ``The Inflationary Universe: A Possible Solution To The Horizon And Flatness
  Problems,''
  Phys.\ Rev.\ D {\bf 23}, 347 (1981).

\bibitem{Albrecht:1982wi}
  A.~Albrecht and P.~J.~Steinhardt,
  ``Cosmology For Grand Unified Theories With Radiatively Induced Symmetry
  Breaking,''
  Phys.\ Rev.\ Lett.\  {\bf 48}, 1220 (1982).

\bibitem{Linde:1982zj}
  A.~D.~Linde,
  ``Coleman-Weinberg Theory And A New Inflationary Universe Scenario,''
  Phys.\ Lett.\ B {\bf 114}, 431 (1982).

\bibitem{Mukhanov:1990me}
  V.~F.~Mukhanov, H.~A.~Feldman and R.~H.~Brandenberger,
  ``Theory of cosmological perturbations. Part 1. Classical perturbations. Part
  2. Quantum theory of perturbations. Part 3. Extensions,''
  Phys.\ Rept.\  {\bf 215}, 203 (1992).

\bibitem{Lyth:1998xn}
  D.~H.~Lyth and A.~Riotto,
  ``Particle physics models of inflation and the cosmological density
  perturbation,''
  Phys.\ Rept.\  {\bf 314}, 1 (1999)
  [arXiv:hep-ph/9807278].

\bibitem{Salopek:1990jq}
  D.~S.~Salopek and J.~R.~Bond,
  ``Nonlinear evolution of long wavelength metric fluctuations in inflationary
  models,''
  Phys.\ Rev.\  D {\bf 42}, 3936 (1990).

\bibitem{Salopek:1990re}
  D.~S.~Salopek and J.~R.~Bond,
  ``Stochastic inflation and nonlinear gravity,''
  Phys.\ Rev.\  D {\bf 43}, 1005 (1991).

\bibitem{Falk:1992sf}
  T.~Falk, R.~Rangarajan and M.~Srednicki,
  ``The Angular dependence of the three point correlation function of the
  cosmic microwave background radiation as predicted by inflationary
  cosmologies,''
  Astrophys.\ J.\  {\bf 403}, L1 (1993)
  [arXiv:astro-ph/9208001].


\bibitem{Gangui:1993tt}
  A.~Gangui, F.~Lucchin, S.~Matarrese and S.~Mollerach,
  ``The Three Point Correlation Function Of The Cosmic Microwave Background In
  Inflationary Models,''
  Astrophys.\ J.\  {\bf 430}, 447 (1994)
  [arXiv:astro-ph/9312033].

\bibitem{Acquaviva:2002ud}
  V.~Acquaviva, N.~Bartolo, S.~Matarrese and A.~Riotto,
  ``Second-order cosmological perturbations from inflation,''
  Nucl.\ Phys.\  B {\bf 667}, 119 (2003)
  [arXiv:astro-ph/0209156].

\bibitem{Maldacena:2002vr}
  J.~M.~Maldacena,
  ``Non-Gaussian features of primordial fluctuations in single field
  inflationary models,''
  JHEP {\bf 0305}, 013 (2003)
  [arXiv:astro-ph/0210603].

\bibitem{Bartolo:2004if}
  N.~Bartolo, E.~Komatsu, S.~Matarrese and A.~Riotto,
  ``Non-Gaussianity from inflation: Theory and observations,''
  Phys.\ Rept.\  {\bf 402}, 103 (2004)
  [arXiv:astro-ph/0406398].

\bibitem{Komatsu:2001rj}
  E.~Komatsu and D.~N.~Spergel,
  ``Acoustic signatures in the primary microwave background bispectrum,''
  Phys.\ Rev.\  D {\bf 63}, 063002 (2001)
  [arXiv:astro-ph/0005036].

\bibitem{Yadav:2007yy}
  A.~P.~S.~Yadav and B.~D.~Wandelt,
  ``Detection of primordial non-Gaussianity (fNL) in the WMAP 3-year data at
  above 99.5\% confidence,''
  arXiv:0712.1148 [astro-ph].

\bibitem{Komatsu:kitpc}
  E.~Komatsu,
  ``$f_{NL}$,''
  talk at String Theory and Cosmology Program in KITPC,
  http://www.kitpc.ac.cn/Activities/program/63574646/fnl.pdf.


\bibitem{Komatsu:2008hk}
  E.~Komatsu {\it et al.}  [WMAP Collaboration],
  ``Five-Year Wilkinson Microwave Anisotropy Probe (WMAP)
  Observations:Cosmological Interpretation,''
  arXiv:0803.0547 [astro-ph].

\bibitem{Linde:1996gt}
  A.~D.~Linde and V.~F.~Mukhanov,
  ``Nongaussian isocurvature perturbations from inflation,''
  Phys.\ Rev.\  D {\bf 56}, 535 (1997)
  [arXiv:astro-ph/9610219].

\bibitem{Enqvist:2001zp}
  K.~Enqvist and M.~S.~Sloth,
  ``Adiabatic CMB perturbations in pre big bang string cosmology,''
  Nucl.\ Phys.\  B {\bf 626}, 395 (2002)
  [arXiv:hep-ph/0109214].

\bibitem{Lyth:2001nq}
  D.~H.~Lyth and D.~Wands,
  ``Generating the curvature perturbation without an inflaton,''
  Phys.\ Lett.\  B {\bf 524}, 5 (2002)
  [arXiv:hep-ph/0110002].

\bibitem{Moroi:2001ct}
  T.~Moroi and T.~Takahashi,
  ``Effects of cosmological moduli fields on cosmic microwave background,''
  Phys.\ Lett.\  B {\bf 522}, 215 (2001)
  [Erratum-ibid.\  B {\bf 539}, 303 (2002)]
  [arXiv:hep-ph/0110096].

\bibitem{Lyth:2002my}
  D.~H.~Lyth, C.~Ungarelli and D.~Wands,
  ``The primordial density perturbation in the curvaton scenario,''
  Phys.\ Rev.\  D {\bf 67}, 023503 (2003)
  [arXiv:astro-ph/0208055].

\bibitem{Bartolo:2003jx}
  N.~Bartolo, S.~Matarrese and A.~Riotto,
  ``On non-Gaussianity in the curvaton scenario,''
  Phys.\ Rev.\  D {\bf 69}, 043503 (2004)
  [arXiv:hep-ph/0309033].



\bibitem{Creminelli:2007aq}
  P.~Creminelli and L.~Senatore,
  ``A smooth bouncing cosmology with scale invariant spectrum,''
  JCAP {\bf 0711}, 010 (2007)
  [arXiv:hep-th/0702165].

\bibitem{Buchbinder:2007at}
  E.~I.~Buchbinder, J.~Khoury and B.~A.~Ovrut,
  ``Non-Gaussianities in New Ekpyrotic Cosmology,''
  arXiv:0710.5172 [hep-th].

\bibitem{Koyama:2007if}
  K.~Koyama, S.~Mizuno, F.~Vernizzi and D.~Wands,
  ``Non-Gaussianities from ekpyrotic collapse with multiple fields,''
  JCAP {\bf 0711}, 024 (2007)
  [arXiv:0708.4321 [hep-th]].

\bibitem{Lehners:2007wc}
  J.~L.~Lehners and P.~J.~Steinhardt,
  ``Non-Gaussian Density Fluctuations from Entropically Generated Curvature
  Perturbations in Ekpyrotic Models,''
  arXiv:0712.3779 [hep-th].





\bibitem{Malik:2002jb}
  K.~A.~Malik, D.~Wands and C.~Ungarelli,
  ``Large-scale curvature and entropy perturbations for multiple interacting
  fluids,''
  Phys.\ Rev.\  D {\bf 67}, 063516 (2003)
  [arXiv:astro-ph/0211602].

\bibitem{Lyth:2006gd}
  D.~H.~Lyth,
  ``Non-gaussianity and cosmic uncertainty in curvaton-type models,''
  JCAP {\bf 0606}, 015 (2006)
  [arXiv:astro-ph/0602285].

\bibitem{Linde:1993xx}
  A.~D.~Linde, D.~A.~Linde and A.~Mezhlumian,
  ``From the Big Bang theory to the theory of a stationary universe,''
  Phys.\ Rev.\  D {\bf 49}, 1783 (1994)
  [arXiv:gr-qc/9306035].


\bibitem{Bunch:1978yq}
  T.~S.~Bunch and P.~C.~W.~Davies,
  ``Quantum Field Theory In De Sitter Space: Renormalization By Point
  Splitting,''
  Proc.\ Roy.\ Soc.\ Lond.\  A {\bf 360} (1978) 117.


\bibitem{Vilenkin:1982wt}
  A.~Vilenkin and L.~H.~Ford,
  ``Gravitational Effects Upon Cosmological Phase Transitions,''
  Phys.\ Rev.\  D {\bf 26}, 1231 (1982).

\bibitem{Linde:1982uu}
  A.~D.~Linde,
  ``Scalar Field Fluctuations In Expanding Universe And The New Inflationary
  Universe Scenario,''
  Phys.\ Lett.\  B {\bf 116}, 335 (1982).




\bibitem{Linde:2005yw}
  A.~Linde and V.~Mukhanov,
  ``The curvaton web,''
  JCAP {\bf 0604}, 009 (2006)
  [arXiv:astro-ph/0511736].

\bibitem{Linde:2005ht}
  A.~D.~Linde,
  ``Particle Physics and Inflationary Cosmology,''
  arXiv:hep-th/0503203.

\bibitem{Linde:2003hc}
  A.~Linde,
  ``Can we have inflation with Omega > 1?,''
  JCAP {\bf 0305}, 002 (2003)
  [arXiv:astro-ph/0303245].

\bibitem{Tegmark:2004qd}
  M.~Tegmark,
  ``What does inflation really predict?,''
  JCAP {\bf 0504}, 001 (2005)
  [arXiv:astro-ph/0410281].


\bibitem{Chun:2004gx}
  E.~J.~Chun, K.~Dimopoulos and D.~Lyth,
  ``Curvaton and QCD axion in supersymmetric theories,''
  Phys.\ Rev.\  D {\bf 70}, 103510 (2004)
  [arXiv:hep-ph/0402059].


\bibitem{Lyth:2003dt}
  D.~H.~Lyth,
  ``Can the curvaton paradigm accommodate a low inflation scale,''
  Phys.\ Lett.\  B {\bf 579}, 239 (2004)
  [arXiv:hep-th/0308110].

\bibitem{Lyth:1995ka}
  D.~H.~Lyth and E.~D.~Stewart,
  ``Thermal Inflation And The Moduli Problem,''
  Phys.\ Rev.\  D {\bf 53}, 1784 (1996)
  [arXiv:hep-ph/9510204].




\bibitem{Huang:2007gk}
  Q.~G.~Huang,
  ``Weak gravity conjecture constraints on inflation,''
  JHEP {\bf 0705}, 096 (2007)
  [arXiv:hep-th/0703071].

\bibitem{Huang:2007qz}
  Q.~G.~Huang,
  ``Constraints on the spectral index for the inflation models in string
  landscape,''
  Phys.\ Rev.\  D {\bf 76}, 061303 (2007)
  [arXiv:0706.2215 [hep-th]].

\bibitem{Lyth:1996im}
  D.~H.~Lyth,
  ``What would we learn by detecting a gravitational wave signal in the  cosmic
  microwave background anisotropy?,''
  Phys.\ Rev.\ Lett.\  {\bf 78}, 1861 (1997)
  [arXiv:hep-ph/9606387].

\bibitem{Dimopoulos:2002kt}
  K.~Dimopoulos and D.~H.~Lyth,
  ``Models of inflation liberated by the curvaton hypothesis,''
  Phys.\ Rev.\  D {\bf 69}, 123509 (2004)
  [arXiv:hep-ph/0209180].

\bibitem{Huang:2007st}
  Q.~G.~Huang,
  ``Weak Gravity Conjecture for the Effective Field Theories with N Species,''
  arXiv:0712.2859 [hep-th].


\bibitem{Dvali:2007wp}
  G.~Dvali and M.~Redi,
  ``Black Hole Bound on the Number of Species and Quantum Gravity at LHC,''
  arXiv:0710.4344 [hep-th].

\bibitem{Freese:2004vs}
  K.~Freese and D.~Spolyar,
  ``Chain inflation: 'Bubble bubble toil and trouble',''
  JCAP {\bf 0507}, 007 (2005)
  [arXiv:hep-ph/0412145].

\bibitem{Huang:2007ek}
  Q.~G.~Huang,
  ``Simplified Chain Inflation,''
  JCAP {\bf 0705}, 009 (2007)
  [arXiv:0704.2835 [hep-th]].


\bibitem{Boyle:2003km}
  L.~A.~Boyle, P.~J.~Steinhardt and N.~Turok,
  ``The cosmic gravitational wave background in a cyclic universe,''
  Phys.\ Rev.\  D {\bf 69}, 127302 (2004)
  [arXiv:hep-th/0307170].



\bibitem{ArkaniHamed:2003uz}
  N.~Arkani-Hamed, P.~Creminelli, S.~Mukohyama and M.~Zaldarriaga,
  ``Ghost inflation,''
  JCAP {\bf 0404}, 001 (2004)
  [arXiv:hep-th/0312100].

\bibitem{Chen:2006nt}
  X.~Chen, M.~x.~Huang, S.~Kachru and G.~Shiu,
  ``Observational signatures and non-Gaussianities of general single field
  inflation,''
  JCAP {\bf 0701}, 002 (2007)
  [arXiv:hep-th/0605045].

\bibitem{ArmendarizPicon:1999rj}
  C.~Armendariz-Picon, T.~Damour and V.~F.~Mukhanov,
  ``k-inflation,''
  Phys.\ Lett.\  B {\bf 458}, 209 (1999)
  [arXiv:hep-th/9904075].

\bibitem{Chen:2007gd}
  B.~Chen, Y.~Wang and W.~Xue,
  ``Inflationary NonGaussianity from Thermal Fluctuations,''
  arXiv:0712.2345 [hep-th].

\bibitem{Li:2007}
  M.~Li, T.~Wang and Y.~Wang,
  ``General Single Field Inflation with Large Positive Non-Gaussianity,''
  arXiv: 0801.0040 [astro-ph].

\bibitem{Dvali:2003em}
  G.~Dvali, A.~Gruzinov and M.~Zaldarriaga,
  ``A new mechanism for generating density perturbations from inflation,''
  Phys.\ Rev.\  D {\bf 69}, 023505 (2004)
  [arXiv:astro-ph/0303591].

\bibitem{Dvali:2003ar}
  G.~Dvali, A.~Gruzinov and M.~Zaldarriaga,
  ``Cosmological perturbations from inhomogeneous reheating, freezeout, and
  mass domination,''
  Phys.\ Rev.\  D {\bf 69}, 083505 (2004)
  [arXiv:astro-ph/0305548].


\bibitem{Matsuda:2007tr}
  T.~Matsuda,
  ``Cosmological perturbations from inhomogeneous preheating and multi-field
  trapping,''
  JHEP {\bf 0707}, 035 (2007)
  [arXiv:0707.0543 [hep-th]].

\bibitem{Matsuda:2007ds}
  T.~Matsuda,
  ``Hybrid Curvatons from Broken Symmetry,''
  JHEP {\bf 0709}, 027 (2007)
  [arXiv:0708.4098 [hep-ph]].

\bibitem{Berera:1995ie}
  A.~Berera,
  ``Warm Inflation,''
  Phys.\ Rev.\ Lett.\  {\bf 75}, 3218 (1995)
  [arXiv:astro-ph/9509049].

\bibitem{Berera:1996nv}
  A.~Berera,
  ``Thermal Properties of an Inflationary Universe,''
  Phys.\ Rev.\  D {\bf 54}, 2519 (1996)
  [arXiv:hep-th/9601134].

\bibitem{Gupta:2002kn}
  S.~Gupta, A.~Berera, A.~F.~Heavens and S.~Matarrese,
  ``Non-Gaussian signatures in the cosmic background radiation from warm
  inflation,''
  Phys.\ Rev.\  D {\bf 66}, 043510 (2002)
  [arXiv:astro-ph/0205152].




\end{thebibliography}
\end{document}